\documentclass{ws-ijmpc}

\begin{document}

\markboth{ F. W. S. Lima}
{Majority-vote model on Opinion-Dependent Networks}
\catchline{}{}{}{}{}

\title{Majority-vote model on Opinion-dependent Network}

\author{ F. W. S. Lima}\footnote{Dietrich Stauffer 
Computational Physics Lab, Departamento de 
F\'{\i}sica,Universidade Federal do Piau\'{\i},
 Teresina, Piau\'{\i}, 64049-550, Brasil}
\address{ Dietrich Stauffer Computational Physics Lab,
 Departamento de 
F\'{\i}sica,Universidade Federal do Piau\'{\i},
Teresina,
Piau\'{\i}, 64049-550, Brasil\footnote{\it Piau\'{\i}, the
Universidade Federal do Piau\'{\i}, Brasil.}\\
fwslima@gmail.com}

\maketitle

\begin{history}

\received{Day Month Year}

\revised{Day Month Year}

\end{history}

\begin{abstract}
We study a nonequilibrium model with up-down symmetry and
 a noise parameter $q$
 known as majority-vote model of M.J. Oliveira $1992$ on 
opinion-dependent network or
Stauffer-Hohnisch-Pittnauer networks.
By Monte Carlo simulations and finite-size scaling relations 
the critical exponents $\beta/\nu$, $\gamma/\nu$, and $1/\nu$
 and points $q_{c}$ and
$U^*$ are obtained. After extensive simulations, we obtain
 $\beta/\nu=0.230(3)$,
 $\gamma/\nu=0.535(2)$, and $1/\nu=0.475(8)$. The calculated 
values of the critical noise parameter and Binder cumulant are
 $q_{c}=0.166(3)$ and $U^*=0.288(3)$. Within the error bars,
 the exponents
obey the relation $2\beta/\nu+\gamma/\nu=1$ and  the 
results presented here demonstrate that the majority-vote model
belongs to a different universality class than the equilibrium 
Ising model on Stauffer-Hohnisch-Pittnauer networks,
 but to the same class as majority-vote 
models on some other networks.


\end{abstract}

\ccode{PACS Nos.:  05.10.Ln; 05.70.Fh; 64.60.Fr;}

\section{Introduction}

The equilibrium Ising model \cite{a3,onsager} has become an
excellent tool to study models of social application 
\cite{latané}. Many of these works are well described in a thorough review
\cite{SPSD}, a more recent summary by Stauffer \cite{stauffer1} and the 
following papers in these special issues on sociophysics in this journal. 
The majority-vote 
model (MVM) of Oliveira \cite{mario} is a nonequilibrium
model of social interaction: individuals of a certain
 population make their decisions based on the opinion
 of the majority of their neighbors. This model has 
been studied for several years by various 
researchers in order to model social and economic systems
\cite{zaklan,zaklan1,limanew,lima1,lima2} in regular 
structures \cite{MVM-regular,lima-malarz,santos,lima3}
 and various other complex
networks \cite{MVM-SW0,MVM-SW1,MVM-ERU,MVM-VD,MVM-ABD,MVM-ABU,MVM-ERD,MVM-APN}.

There are applications to real elections in which similar models of 
opinion dynamics have been explored in the literature, 
such as \cite{TVP}.

In the present work, we study the critical properties 
of MVM on Stauffer-Hohnisch-Pittnauer (SHP) networks.
 Hohnisch bonds of SHP networks \cite{LHS,TH,DS}
are links connecting nodes with different values (spins,
 opinions, etc.) on them; they
are at each time step with a low probability $0.0001$ 
replaced by a link to another
randomly selected node. Links connecting agreeing 
nodes are not replaced. In the
present work, we start with each node having 
links to four randomly selected
neighbors. Thus our SHP networks are similar 
to Small-World (Watts-Strogatz)
networks but start from a random network instead of a 
square lattice and use opinion-dependent
(instead of random) rewiring. All links are directed\cite{LHS,TH}. 
The critical exponents and noise parameter were obtained 
using Monte Carlo simulation (MC) and with a finite size
scaling analysis. The
effective dimension of the SHP network is also determined
 for MVM. Finally, the critical exponents
calculated for SHP networks are compared with the
 results obtained for {\it undirected} and {\it directed} 
Barab\'asi-Albert networks (UBA and DBA) \cite{MVM-ABD,MVM-ABU}
 and Erd\"os-R\`enyi random
graphs (UER and DER) \cite{MVM-ERU,MVM-ERD}. 
\section{Model and simulation}

Our network is SHP type composed of $N$ sites and $k=4$ neighbors.
On the MVM model, the system dynamics is as follows. 
Initially, we assign a spin variable
$\sigma$ with values $\pm 1$ at each node of the network. At each step
we try to spin flip a node. The flip is  accepted with
 probability 
\begin{equation} 
w_i=\frac{1}{2}\left[ 1-(1-2q)\sigma_{i}\cdot\text{S}
\left(\sum_{j=1}^k\sigma_j\right)\right],
\label{eq_1}
\end{equation}
where $S(x)$ is the sign $\pm 1$ of $x$ if $x\neq0$, $S(x)=0$ if
$x=0$. To calculate $w_i$ our sum runs over the $k$ nearest
neighbors of spin $i$.  Eq.~(\ref{eq_1}) means that
with probability $(1-q)$ the spin will adopt the same 
state as the majority
of its neighbors. 

Here, the control parameter $0\le q\le 1$
plays a role similar to the temperature in equilibrium 
systems: the smaller
$q$, the greater the probability of parallel aligning 
with the local majority.

To study the critical behavior of the model we define
 the variable $m\equiv\sum_{i=1}^{N}\sigma_{i}/N$.
In particular, we are interested in the magnetization
 $M$, susceptibility $\chi$ and the reduced fourth-order
 cumulant $U$
\begin{subequations}
\label{eq-def}
\begin{equation}
M_{N}(q)\equiv \langle|m|\rangle,
\end{equation}
\begin{equation}
\chi_{N}(q)\equiv N\left(\langle m^2\rangle-\langle m \rangle^2\right),
\end{equation}
\begin{equation}
U_{N}(q)\equiv 1-\dfrac{\langle m^{4}\rangle}{3\langle m^2 \rangle^2},
\end{equation}
\end{subequations}
where $\langle\cdots\rangle$ stands for a thermodynamics average.

The results are averaged over the $N_{\text{run}}$ 
independent simulations. 
These quantities are functions of the noise parameter
 $q$ and obey the finite-size scaling relations
\begin{subequations}
\label{eq-scal}
\begin{equation}
\label{eq-scal-M}
M_{N}(q)=N^{-\beta/\nu}f_m(x),
\end{equation}
\begin{equation}
\label{eq-scal-chi}
\chi_{N}(q)=N^{\gamma/\nu}f_\chi(x),
\end{equation}
\begin{equation}
\label{eq-scal-dUdq}
\frac{dU_{N}(q)}{dq}=N^{1/\nu}f_U(x),
\end{equation}
where $\nu$, $\beta$, and $\gamma$ are the usual critical 
exponents, $f_{m,\chi,U}(x)$ are the finite size scaling functions with
\begin{equation}
\label{eq-scal-x}
x=(q-q_c)N^{1/\nu}
\end{equation}
\end{subequations}
being the scaling variable. Therefore, from the size 
dependence of $M$ and $\chi$ we obtained the exponents
$\beta/\nu$ and $\gamma/\nu$, respectively. 
The maximum value of susceptibility 
also scales as $N^{\gamma/\nu}$. Moreover, the value
 of $q^*$ for which $\chi$ has a maximum is expected 
to scale with the system size $N$ as
\begin{equation}
\label{eq-q-max}
q^*=q_c+bN^{-1/\nu} \text{ with } b\approx 1.
\end{equation}

Therefore, the relations \eqref{eq-scal-dUdq} and
 \eqref{eq-q-max} may be used to get the exponent
$1/\nu$. We also have applied 
the calculated exponents to the 
hyperscaling hypothesis
\begin{equation}
\label{eq-q-def}
2\beta/\nu + \gamma/\nu=D_{eff}
\end{equation}
in order to get the effective dimensionality, $D_{eff}$, 
for connectivity $k$.

We performed Monte Carlo simulation on the SHP networks
 with various systems 
sizes $N$ (250, 500, 1000, 2000, 4000, 8000 and
 16000 sites).
It takes $2\times 10^5$ Monte Carlo steps (MCS) to 
make the system reach the steady state, and then 
the time averages are estimated over the next $2\times 10^5$ MCS.
One MCS is accomplished after all the $N$ spins are 
investigated whether they flip or not.

The results are averaged over $N_{\text{run}}$ $(50\le N_{\text{run}} \le 100)$ independent 
simulation runs for each network and for given set 
of parameters $(q,N)$. Here, were used $50$ independent networks
for each system size $N$ cited earlier.
\begin{table}[ht]
\tbl{ The critical noise $q_{c}$, the critical exponents,
and the effective dimensionality 
$D_{eff}$, for  DBA, UBA, DER, UER, and SHP network with 
connectivity $k=4$. Error bars are statistical only. ${\gamma/\nu}^{q_{c}}$
is calculated from $\chi$ at $q_c$ and ${\gamma/\nu}^{q_{c}(N)}$ from the
maximal $\chi$.}
{\begin{tabular}{|c c c c c c c c|}
\hline
\hline
$ k=4 $ & $q_{c}$ & $\beta/\nu$& ${\gamma/\nu}^{q_{c}}$ & ${\gamma/\nu}^{q_{c}(N)}$ & $ 1/\nu$ & $ D_{eff}$ & Ref.\\
\hline
$DBA $ & $ 0.431(3) $ & $ 0.447(2) $ & $ 0.856(15)$ & $ 0.888(9) $ & $ - $ & $ 0.998(3)$ & \cite{MVM-ABD}\\
$UBA $ & $ 0.306(3) $ & $ 0.231(22)$ & $ 0.537(8) $ & $ 0.519(17) $ & $ 0.43(2) $ & $ 0.999(23)$ & \cite{MVM-ABU}\\
$DER $ & $ 0.175(4) $ & $ 0.230(5) $ & $ 0.530(6) $ & $ 0.516(2) $ & $ 0.545(26)$ & $ 0.990(7)$ & \cite{MVM-ERD}\\
$UER $ & $ 0.181(1) $ & $ 0.242(6) $ & $ 0.54(1) $  & $ 0.515(6) $ & $ 0.59(7) $ & $ 1.02(2)$ & \cite{MVM-ERU}\\
$SHP $ & $ 0.166(3) $ & $ 0.230(3) $ & $ 0.535(2)$  & $ 0.523(5) $ & $ 0.475(8) $ & $ 0.995(3)$ & here\\
\hline
\hline
\end{tabular} \label{table1}}
\end{table}
\section{Results and Discussion}

\begin{figure}[ph]
\centerline{\psfig{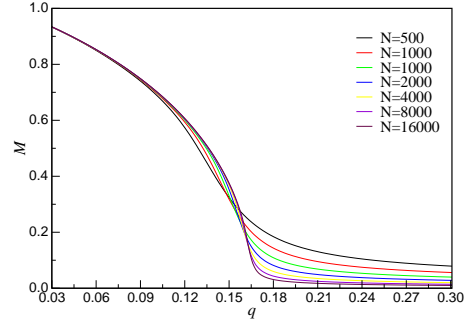}}
\vspace*{8pt}
\caption{Plot of the magnetization $M$ as a function of 
the noise parameter $q$, for $N=250$, $500$, $1000$,
 $2000$, $4000$, $8000$, and $16000$ sites. \label{f1}}
\end{figure}

\begin{figure}[ph]
\centerline{\psfig{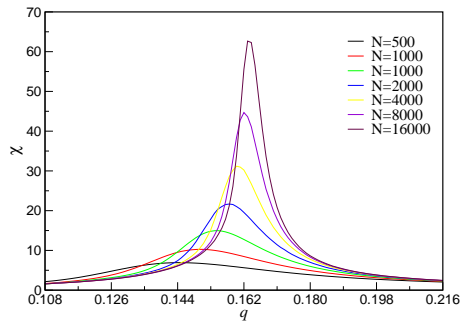}}
\vspace*{8pt}
\caption{The same Fig.~\ref{f1}, but now for the 
susceptibility $\chi$ as a function of the noise 
parameter $q$. \label{f2}}
\end{figure}

\begin{figure}[ph]
\centerline{\psfig{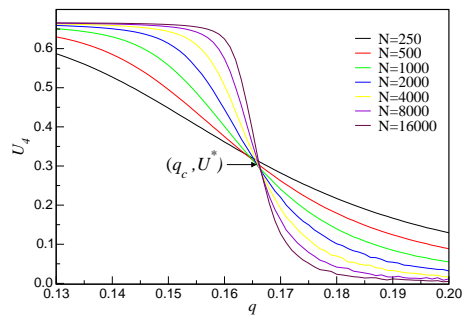}}
\vspace*{8pt}
\caption{The same the Fig.~\ref{f1}, but now for Binder
 cumulant $U$ as a function of the noise parameter $q$. \label{f3}}
\end{figure}

In Figs.~\ref{f1}, ~\ref{f2}, and ~\ref{f3}  we show the 
dependence of the magnetization $M$, susceptibility $\chi$,
 and Binder cumulant $U$ on the noise parameter $q$, 
obtained from simulations on $SHP$
 networks with $L$ ranging from $N=250$ to $16000$ sites. 
The shape of $M(q)$, $\chi(q)$, and
 $U(q)$ curve, for a given value of $N$, suggests the
 presence of a second-order phase transition in the 
system. The phase transition occurs at the critical value $q_c$ of the
 noise parameter $q$.
This parameter $q_c$ is estimated as the
 point where the $U_N(q)$ curves for different system sizes $N$ 
intercept each other \cite{binder}.
Then, we obtain $q_c=0.166(3)$ and $U^*=0.288(3)$ for
 $SHP$ networks.
\bigskip
\begin{figure}[ph]
\centerline{\psfig{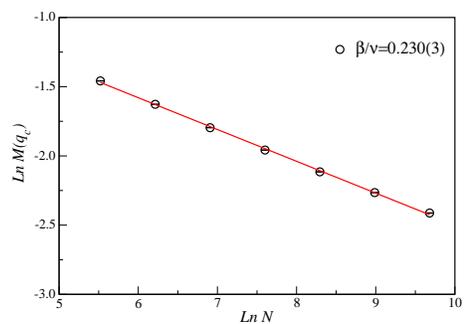}}
\vspace*{8pt}
\caption{ Plot of the magnetization $M^*=M(q_c)$ vs. the
 linear system size $N$ SHP network. \label{f4}}
\end{figure}
\begin{figure}[ph]
\centerline{\psfig{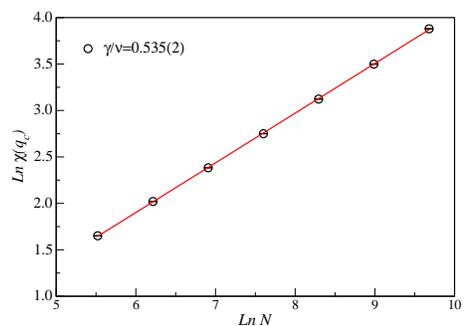}}
\vspace*{8pt}
\caption{Display of the susceptibility at $q_c$ versus 
$N$ for SHP network. \label{f5}}
\end{figure}
\begin{figure}[ph]
\centerline{\psfig{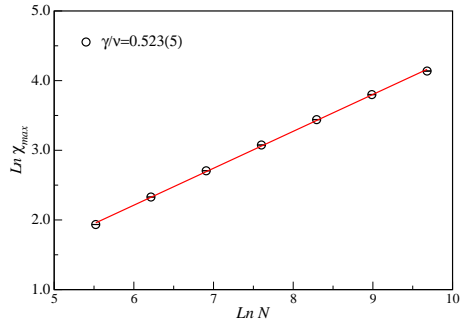}}
\vspace*{8pt}
\caption{Plot of the susceptibility at $q_{\chi_{max}}(N)$ 
versus $N$ for for SHP network. \label{f6}}
\end{figure}
\begin{figure}[ph]
\centerline{\psfig{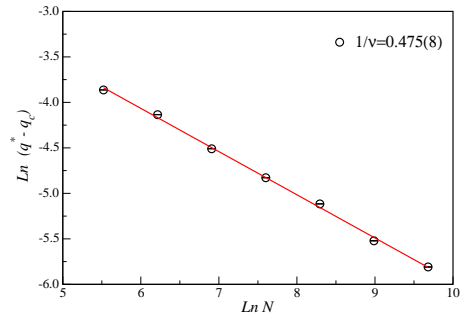}}
\vspace*{8pt}
\caption{Plot $\ln|q_c(N)-q_c|$ versus the system size 
$N$ for SHP network. \label{f7}}
\end{figure}
\begin{figure}[ph]
\centerline{\psfig{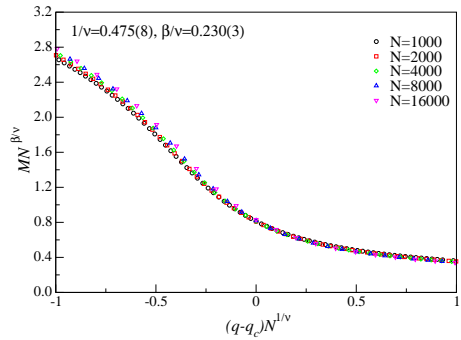}}
\vspace*{8pt}
\caption{Plot of the data collapse of the magnetisation 
{\it M} for the  system size $N=1000$,$2000$,
$4000$,$8000$, and $16000$ for $SHP$ network. The exponents 
used here were
$\beta/\nu=0.230(3)$ and $1/\nu=0.475(8)$. \label{f8}}
\end{figure}
\begin{figure}[ph]
\centerline{\psfig{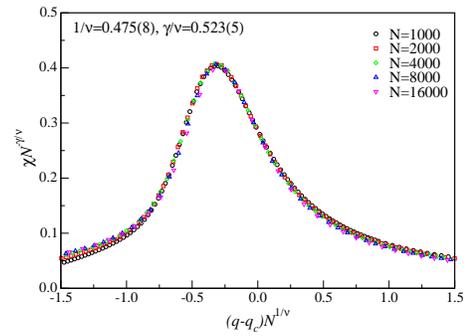}}
\vspace*{8pt}
\caption{Data colapse of the susceptibility for the  system 
size $N=1000$,$2000$,
$4000$,$8000$, and $16000$ for $SHP$ network. The exponents 
used here were
$\gamma/\nu=0.523(5)$ and $1/\nu=0.475(8)$. \label{f9}}
\end{figure}

In  Fig.~\ref{f4} we plot the dependence of the magnetization 
$M^*=M(q_c)$ vs. the system size $N$.
The slope of curve corresponds to the exponent ratio $\beta/\nu$
 according to Eq. (3a).
The obtained exponent is $\beta/\nu= 0.230(3)$ for our $SHP$
 network.

The exponent ratio $\gamma/\nu$ at $q_{c}$ is obtained from 
the slope of the straight line with $\gamma/\nu= 0.535(2)$ , 
as presented in Fig. \ref{f5}. The exponents ratio $\gamma/\nu$
 at $q_{\chi_{max}}(N)$ 
is $\gamma/\nu=0.523(5)$ for $SHP$ networks as presented in Fig. \ref{f6}.

To obtain the critical exponent $1/\nu$, we used the scaling relation (4).
The calculated  value of the exponent $1/\nu$ are $1/\nu=0.475(8)$ 
for SHP networks (see Fig. \ref{f7}).
We plot $MN^{\beta/\nu}$ versus $(q-q_{c})N^{1/\nu}$ in Fig.
 \ref{f8} using the critical
exponents $1/\nu=0.475(8)$ and $\beta/\nu=0.230(3)$ for system 
size $N=1000$,$2000$,
$4000$,$8000$, and $16000$ for $SHP$ network. The excellent collapse of 
the curves for five different system sizes corroborates the estimate 
for $q_c$ and the critical exponents $\beta/\nu$ and $1/\nu$.

In Fig. \ref{f9} we plot $\chi N^{-\gamma/\nu}$ 
versus $(q-q_{c})N^{1/\nu}$
 using the critical exponents $\gamma/\nu=0.523(5)$ and $1/\nu=0.475(8)$ 
for system size $N=1000$, 2000,
4000, 8000, and 16000 for $SHP$ network. Again, the excellent collapse of 
the curves for five different system size corroborates the extimation
 for $q_c$ and the critical exponents $\gamma/\nu$ and 
$1/\nu$. The results of simulations are collected in Tab. \ref{table1}.
\section{Conclusion} 

The determination of the universality class of the MVM model on 
differents non-regular structure as Small-Worlds, scale-free networks, 
random graphs, and others has been studied by many 
researchers in recent years 
\cite{MVM-SW0,MVM-SW1,MVM-ERU,MVM-VD,MVM-ABD,MVM-ABU,MVM-ERD,MVM-APN}. 
Finally, here, we remark that our MC results obtained on $SHP$ network for
 MVM model show that critical exponent ratios $\beta/\nu$ and $\gamma/\nu$
 are similar to those for $UBA$ networks, $UER$ and $DER$ random graphs 
for values of conectivity $k=4$, 
but different from the results of $DBA$ networks and MVM 
model for regular
 lattice \cite{mario} and equilibrium 2D Ising model 
\cite{onsager}. Here, 
we show also that the critical exponent $1/\nu$ is different  
from random Erd\"os-R\'enyi graphs. Therefore, unfortunately, 
because of the critical 
exponent ratio $1/\nu$, we cannot assert that
MVM models
 on different 
structures as scale-free ($UBA$) network and random graphs 
($UER$ and $DER$) 
belong to the same universality class of
the MVM model on $SHP$ network; only the critical exponent
ratios 
$\beta/\nu$ and $\gamma/\nu$ are similar to $UBA$ networks and to 
 $UER$ and $DER$
random graphs. Here, we also showed that the effective dimension 
$D_{eff}$ is close 
to $1$ for all networks and graphs studied in this work.
The agreement in $D_{eff}$ and the two exponent ratios but not in $1/\nu$
(weak universality) remains to be explained. 
\bigskip

\section*{Acknowledgments}
The author thanks D. Stauffer for many suggestion and 
fruitful discussions
 during the
development this work and also for reading this paper. 
We also acknowledge the
Brazilian agency CNPQ for  its financial support. This
work also was supported the system SGI Altix 1350 in the computational park
CENAPAD.UNICAMP-USP, SP-BRAZIL.

\end{document}